# Corrigendum to: A Systematic Study of DDR4 DRAM Faults in the Field

Majed Valad Beigi[1], Yi Cao[2], Sudhanva Gurumurthi[3], Charles Recchia[4,a], Andrew Walton[5,b], Vilas Sridharan[1]

[1] RAS Architecture, Advanced Micro Devices, Inc., Boxborough, MA. Emails: majed.valadbeigi@amd.com and vilas.sridharan@amd.com.
[2] Google, Sunnyvale, CA. Emails: smartcaoyi@google.com.
[3] RAS Architecture, Advanced Micro Devices, Inc., Austin, TX. Email: sudhanva.gurumurthi@amd.com
[4] ASMPT NEXX, Inc., Boston, MA, Austin, TX. Email: charles.recchia@ieee.org.
[5] Microsoft, Mountain View, CA. Email: dwalton64@gmail.com.

## ABSTRACT

This paper is a corrigendum to the paper by Beigi et al. published at HPCA 2023 https://doi.org/10.1109/HPCA56546.2023.10071066. The HPCA paper presented a detailed field data analysis of faults observed at scale in DDR4 DRAM from two different memory vendors. This analysis included a breakdown of fault patterns or modes. Upon further study of the data, we found a bug in how we decoded errors based on the logged row-bank-column address. Specifically, we found that some errors that occurred in one column were mis-interpreted as occurring in two non-adjacent columns. As a result of this, some single-bit faults were misclassified as partial-row faults (i.e., two-bit faults). Similarly, some single-column faults were misclassified as two-column faults. The result of these misclassification errors is that the proportion of single-bit faults is higher than reported in the paper, with a commensurate reduction in the fraction of certain types of multi-bit faults. These misclassifications also slightly change the Failure In Time (FIT) per DRAM device values presented in the original paper.

In this corrigendum, we provide an updated version of the relevant tables and figures and point out the corresponding page numbers and references in the original paper that they replace.

## CORRIGENDUM

This corrigendum is prepared to provide the updated tables, figures, and sentences in some sections of the original paper (i.e., the HPCA 2023 paper referenced in the abstract). In the following, we point to each section from the original paper and provide the updated tables, figures, and sentences. Note that the page numbers provided correspond to those in the HPCA proceedings that the DOI URL referenced in the abstract.

### *Abstract*

In page 991 of the original paper, line 6 in the second paragraph of the abstract is changed to "over 23% of the faults that occurred affected multiple DRAM bits."

### I. INTRODUCTION

In Section I of the original paper, lines 3 and 4 of the first bullet in page 992 are changed to "*intermittent fault rates vary by 1.43x; and permanent fault rates vary by 2.01x*". Moreover, the text of the second bullet in page 992 is changed to *"~23% of the faults that occur in the DDR4 DRAM system under study affect multiple bits, such as faults that affect all or part of a DRAM row, column, or bank*". Note that lines 4 to 7 in the second bullet are also removed since the updated percentage of multiple faults occurred in the DDR4 is close to those described in references [2] and [11] of the paper.

### VII. DRAM, DIMM, AND CHANNEL FAULTS

#### *A. Fault Rates by Vendor*

---

[a] This work was done when the author was at AMD
[b] This work was done when the author was at Google

In the updated version of the paper, the FIT/DRAM device of intermittent and permanent faults for both vendors are slightly changed. Hence, we updated Figure 2 in Page 994. Moreover, in page 995, the last line of the first paragraph in Section VII.A is changed to "*vendor A by 1.43x and 2.01x, respectively*".

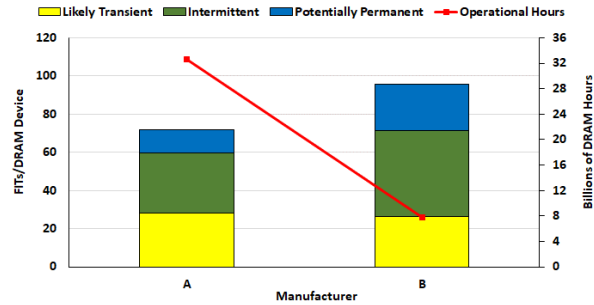

Figure 2. Fault rates (left axis) and operational hours (right axis) by Vendor.

TABLE I. THE PERCENTAGE OF FAULT MODES AND TYPES.

| Fault Modes | Total Faults | | Fault Type | | | | | |
|---|---|---|---|---|---|---|---|---|
| | Vendor A | Vendor B | Vendor A | | | Vendor B | | |
| | | | Likely Transient | Intermittent | Potentially Permanent | Likely Transient | Intermittent | Potentially Permanent |
| Single-bit | 76.74% | 77.27% | 32.83% | 50.53% | 16.64% | 31.91% | 47.16% | 20.93% |
| Single-word | 1.38% | 0.00% | 46.88% | 50.00% | 3.12% | 0% | 0% | 0% |
| Single-column | 6.56% | 4.79% | 75.66% | 17.76% | 6.58% | 20.00% | 54.29% | 25.71% |
| Two-column | 5.65% | 0.14% | 95.42% | 3.82% | 0.76% | 0.00% | 0.00% | 100% |
| Partial-row | 2.89% | 1.23% | 4.48% | 73.13% | 22.39% | 11.11% | 77.78% | 11.11% |
| Single-row | 0.52% | 0% | 0% | 50.00% | 50.00% | 0% | 0% | 0% |
| Single-row-plus-single-bit | 0% | 1.10% | 0% | 0% | 0% | 37.50% | 50.00% | 12.50% |
| Two-row | 0.13% | 7.81% | 33.33% | 33.33% | 33.34% | 1.75% | 54.39% | 43.86% |
| Consecutive-row | 0.43% | 0.68% | 60.00% | 0.00% | 40.00% | 0% | 80.00% | 20.00% |
| Cluster-row | 4.10% | 4.66% | 60.00% | 11.58% | 28.42% | 23.53% | 29.41% | 47.06% |
| Single-bank | 0.13% | 0% | 0% | 0% | 100% | 0% | 0% | 0% |
| Quarter-device | 0% | 0.27% | 0% | 0% | 0% | 0% | 50.00% | 50.00% |
| Half-device | 0.04% | 0.14% | 0% | 0% | 100% | 0% | 0% | 100% |
| Full-device | 0.52% | 0.68% | 41.67% | 16.67% | 41.66% | 0% | 60.00% | 40.00% |
| Single-pin | 0.52% | 0.82% | 0% | 8.33% | 91.67% | 16.67% | 16.67% | 66.66% |
| Single-lane | 0.39% | 0.41% | 22.22% | 22.22% | 55.56% | 0.00% | 0.00% | 100.00% |
| Total | 100% | 100% | 39.39% | 43.96% | 16.65% | 27.53% | 47.4% | 25.07% |

*B. Fault Modes*

In this section, we updated the percentage of "Total Faults" and "Fault Type" for some of the fault modes in Table I in page 998 of the original paper. In addition, we modified lines 3 and 4 of the first paragraph in page 998 and added the updated percentage of single-bit and multiple bit faults. Specifically, we updated the sentences to "*The table shows that for both vendors ~77% of all DRAM faults are single-bit faults, while the remaining ~23% of all DRAM faults affect multiple bits*". Moreover, the third paragraph in page 998 of the original paper is removed because of the change in the percentage of partial-row faults. Finally, the percentage of bank faults indicated in the last two lines of the fourth paragraph in page 998 is changed to 10.44% and 14.39% for vendor A and vendor B, respectively.

The FIT per DRAM device for some of the fault modes mentioned in Table II in page 998 of the original paper is also updated. Lines 4 to 6 of the first paragraph in Page 999 are also changed to "*However, for single-word, single-column, two-column, partial-row, single-row and single-bank, vendor A has a higher fault rate than vendor B*."

According to Table II, the overall FIT per DRAM device for vendor A and vendor B are still the same as the values that were reported in the original paper. Moreover, the updated FIT per DRAM device of single-bit faults for DDR4 is higher than the one reported for DDR3 [11]. While we could not root cause the precise reason for this trend, our observations are consistent with expectations from the memory industry that DRAM scaling is likely to cause a higher rate of single-bit faults and the introduction of Single-Error Correct In-DRAM ECC in DDR5 to address such faults [38].

TABLE II. THE FAULT RATE (FIT PER DRAM DEVICE) OF FAULT MODES

| Fault Modes | FIT per DRAM Device | |
| --- | --- | --- |
| | Vendor A | Vendor B |
| Single-bit | 54.92 | 73.5 |
| Single-word | 1.03 | 0.02 |
| Single-column | 4.76 | 4.75 |
| Two-column | 4.11 | 0.2 |
| Partial-row | 2.15 | 1.3 |
| Single-row | 0.4 | 0.02 |
| Single-row-plus-single-bit | 0.01 | 1.2 |
| Two-row | 0.11 | 7.6 |
| Consecutive-row | 0.33 | 0.72 |
| Cluster-row | 3.0 | 4.6 |
| Single-bank | 0.11 | 0.02 |
| Quarter-device | 0.01 | 0.31 |
| Half-device | 0.04 | 0.17 |
| Full-device | 0.4 | 0.72 |
| Single-pin | 0.4 | 0.86 |
| Single-lane | 0.3 | 0.45 |
| Total | 72.08 | 96.44 |

*D. Fault Types*

The ratios of intermittent faults to likely transient faults and intermittent faults to potentially permanent faults indicated in lines 3 to 7 of the paragraph in Section VII.D in Page 999 are updated. Specifically, the sentences are changed to "*The table shows that the ratio of intermittent faults to likely transient faults is 1.71x for vendor B and 1.12x for vendor A, while the ratio of intermittent faults to potentially permanent faults is 1.88x for vendor B and 2.63x for Vendor A*".

The updated results in Table I show that the percentage of intermittent faults for single-bit faults is higher than the percentage of transient faults for both vendors. Hence, single-bit faults are intermittent for both vendors. Also, single-word faults are approximately equally likely to be transient or intermittent for vendor A.

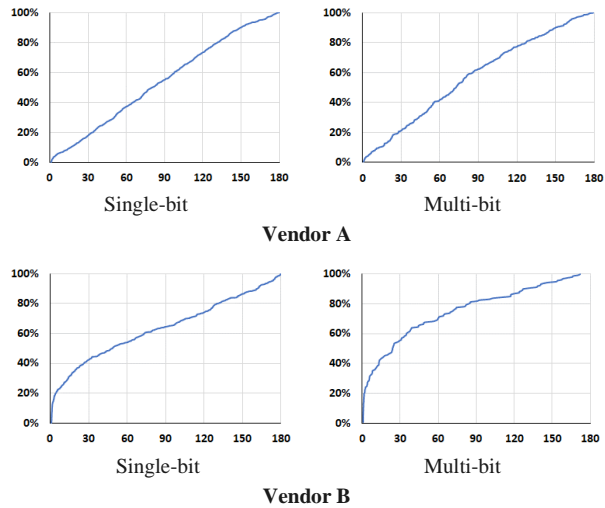

Figure 6. Time to fail in days. Single-bit and multi-bit faults for vendor A show a linear TTF, indicating a constant failure rate, while single-bit and multi-bit faults for Vendor B show a high rate of early-life failures followed by a lower steady-state rate.

*E. Time To Fail*

We updated Figure 6 in Page 999 of the original paper that shows time to fail of single-bit and multi-bit faults for both vendors.

VIII. CASE STUDY: COMPARING ERROR-CORRECTING CODES

In this section, the percentage of faults that affect multiple bits mentioned in line 10 of the first paragraph in page 1000 is changed to "~23%". In addition, in line 14 of the first paragraph of Section VIII in page 1000, we updated the rate of faults that cause uncorrectable errors (i.e., UE Fault) when using SEC-DED ECC to 17.16 and 22.94 FIT per DRAM device for vendor A and vendor B, respectively.

*C. Results*

In Table V located in page 1001 of the original paper, the rate of faults that cause uncorrectable errors (i.e., UE Faults) when using SEC-DED ECC is updated for both vendor A and vendor B.

Note that the misclassification explained in the abstract of the paper only affects the FIT per DRAM device of cluster-row and two-column faults with 1DQ. As a result, the UE faults rate of bounded fault ECC for vendor A and vendor B are the same as the values that were reported in the original paper.

The updated results in Table V show that the UE Faults rate when using SEC-DED ECC is still higher than the UE faults rate when using bounded fault ECC (~3X for vendor A and ~5X for vendor B). These results confirm that we still need a stronger ECC scheme (i.e., Chipkill) specifically when the amount of installed DRAM at hyperscale scales by orders of magnitude.

TABLE V. RATE OF FAULTS IN A SINGLE DRAM DEVICE THAT CAUSE UNCORRECTABLE ERRORS (I.E., UE FAULTS)

| DIMM Configs | ECC | FIT/DRAM Device | |
|---|---|---|---|
| | | Vendor A | Vendor B |
| DDR4 (18×4) | Chipkill | 0 | 0 |
| | SEC-DED | 17.16 | 22.94 |
| | Bounded Fault | 5.1 | 4.44 |
| DDR5 (10×4) | Chipkill | 0 | 0 |
| | Bounded Fault | 5.71 | 4.97 |
| DDR5 (9×4) | Bounded Fault | 5.35 | 4.58 |

IX. CONCLUSION

In line 2 of the second paragraph of the conclusion section in Page 1001, the percentage of all faults in our DDR4 data set that affect multiple bits is changed to "~23%"

## ACKNOWLEDGEMENTS

© 2024 Advanced Micro Devices, Inc. All rights reserved. AMD, the AMD Arrow logo, and combinations thereof are trademarks of Advanced Micro Devices, Inc. Other product names used in this publication are for identification purposes only and may be trademarks of their respective companies. This paper reflects collaborative work between the authors. This paper does not express nor imply an official statement or any viewpoints of Advanced Micro Devices, Inc. (AMD).